\documentclass[%
 reprint,
 amsmath,amssymb,
 aps,
]{revtex4-2}

\usepackage{xcolor}
\usepackage{lipsum}
\usepackage{braket}
\usepackage{graphicx}
\usepackage{dcolumn}
\usepackage{bm}

\begin{document}

\title{Three Lagrangians for the complete-active space coupled-cluster method}

\author{Simen Kvaal}
\email{simen.kvaal@kjemi.uio.no}
\affiliation{Hylleraas Centre for Quantum Molecular Sciences, Department of Chemistry, University of Oslo, P.O. Box 1033 Blindern, N-0315 Oslo, Norway}%

\date{\today}

\begin{abstract}
Three fully variational formulations of the complete-active space coupled-cluster (CASCC) method are derived. The formulations include the ability to approximate the model vectors by smooth manifolds, thereby opening up the possibility for overcoming the exponential wall of scaling for model spaces of CAS type. In particular, model vectors of matrix-product states are considered, and it is argued that the present variational formulation allows not only favorably-scaling multireference coupled-cluster calculations, but also systematic correction of tailored coupled-cluster calculation and of quantum chemical density-matrix renormalization group methods, which are fast and polynomial scaling, but lacks the ability to properly resolve dynamical correlation at chemical accuracy. The extension of the variational formulations to the time-domain is also discussed, with derivations of abstract evolution equations.  
\end{abstract}

\maketitle


\newcommand{\note}[1]{{\color{red}#1}}

\section{Introduction}
\label{sec:introduction}

In this article, we discuss some Lagrangian formulations of the complete-active space coupled-cluster (CASCC) method introduced in the 1990s by Oliphant, Adamowicz and Piecuch~\cite{Oliphant1991,Oliphant1992,Piecuch1993}, see also the work by Stolarczyk~\cite{Stolarczyk1994}. For a comprehensive review with literature, see Ref.~\cite{Lyakh2012}. More specifically, we will consider the \emph{semilinear} wavefunction formulation of Adamowicz and coworkers~\cite{Adamowicz1996,Adamowicz2000} but \emph{linked} energy-independent working equations to the best of our knowledge first proposed rather recently by Kowalski~\cite{Kowalski2018}. 

The motivation for the present work is threefold: Firstly, any multireference approach employing a complete model space (i.e., full CASCI model vectors) face an exponential wall of scaling for large systems. Systems that are truly strongly correlated requires CASs with the number of orbitals proportional to the number of particles, which is clearly not feasible. A formulation of CASCC that allows approximations of the CASCI model vector in terms of differentiable manifolds such as matrix-product states (MPS) and more generally using the density-matrix renormalization group algorithm (DMRG)~\cite{White1999,Legeza2003,Chan2004,Chan2005,Wouters2014} could enable affordable and accurate multireference calculations for quite large systems. The MPS ansatz is excellent at resolving static correlation at polynomial cost, and there is a growing interest in its application in quantum chemistry. Key to achieving a successful combination of coupled-cluster methods with  such approximations is a fully variational framework. Secondly, literature searches for fully variational formulations of the semilinear CASCC method similar to the formulation which has now become standard in traditional single-reference coupled-cluster (SRCC) theory have been fruitless. The Lagrangian formulation of SRCC was the culmination of several independent researchers' efforts in the 1970--1980s to provide easily computable molecular properties and response to external perturbations~\cite{Monkhorst1977,Arponen1983,Adamowicz1984,Fitzgerald1986,Helgaker1988,Salter1989,Helgaker1989}. 

Although the standard Lagrangian can be used for CASCC when the linear CAS amplitudes are mapped to an exponential ansatz, see e.g.~\cite{Lyakh2012}, a Lagrangian for the semilinear case is desired, since the mapping from linear to exponential parameterizations can be cumbersome and numerically ill-conditioned, especially when excited states are targeted. Moreover, the numerical task of solving a linear (non-Hermitian) eigenvalue problem is different from solving a polynomial root equation.

Thirdly, a fully variational formulation of CASCC would allow its use in real-time propagation of coupled-cluster quantum states, a topic which has garnered recent interest in the attophysics community~\cite{Kvaal2012,Sato2018,Pathak2020,Pathak2020a,Hansen2019,Pedersen2019,Ofstad2023}. As the MPS ansatz is also being employed for real-time propagation of quantum chemical systems~\cite{Lubich2015,Frahm2019,Baiardi2019}, one can speculate that a time-dependent CASCC method with a MPS-based model space description can be far superior to existing approaches.

The CASCC method is well-known to suffer from some issues that originate with its single-reference formalism: The use of a formal reference in the CAS gives a bias towards this reference. This in turn leads to undesirable potential-surface discontinuities. On the other hand, CASCC is relatively straight-forward to implement and to analyze compared to true multireference CC approaches. Moreover, when absolute energies are compared, CASCC seems to have similar accuracy as the genuine MRCC approches~\cite{Bodenstein2020}. A closely related method is the tailored coupled-cluster method (TCC)~\cite{Kinoshita2005}, an externally corrected CC approach that is actively being studied, including the DMRG-TCC method where the CAS model function is replaced by a matrix-product state and optimized with the DMRG algorithm\cite{Veis2016,Faulstich2019,Faulstich2019b,Lang2020,Morchen2020}. The TCC method, however, suffers from a systematic error due to the CAS wavefunction being computed once and for all, and not being computed self-consistently with the external correction. The present work could therefore pave the way for updating schemes for the matrix-product state that reduce or altogether eliminate such systematic errors.

We will derive three fully variational formulations of CASCC, which we call Lagrangians even if some additional variables introduced are not strictly speaking Lagrange multipliers. After a brief discussion of the CASCC method in Section~\ref{sec:cascc}, the first Lagrangian is introduced in Section~\ref{sec:lag1}. The Lagrangian has a similar interpretation as the standard SRCC Lagrangian as a constrained optimization of the CC energy. This Lagrangian is applicable to standard CASCC calculations where the full CASCI model vector is used, and where the linked form of the amplitude equations is used. (These are equivalent to the unlinked amplitude equations in just about any practical truncation scheme.) Just as the SRCC Lagrangian features a Lagrange multiplier operator that supplements the exponential ansatz, the CASCC Lagrangian features, in a very natural manner, a ``Lagrange multiplier'' CASCI model vector, i.e., we have \emph{two} model vectors in addition to \emph{two} cluster operators. The working equations for the model vectors are a right-eigenvalue problem for the CAS projection of the connected Hamiltonian, a non-Hermitian operator, yielding the CASCI model vector as solution, and an \emph{inhomogenous} left-eigenvalue problem, giving the dual CASCI model vector as solution. The asymmetry of the rôles of these vectors can be traced to the ``ket-centric'' method of development of CASCC as a projection technique. 

In Section~\ref{sec:variational} we introduce a differenetiable submanifold of the CAS in which to approximate the CASCI model vectors. (This manifold must not be confused with what some authors call ``the projection manifold'' in CC theory.) Thus, both model vectors are assumed to lie on the manifold, and the corresponding variational equations are derived. However, the asymmetry of the equations for the two model vectors implies that available codes for the variational approximation of non-Hermitian eigenvalue problems cannot be reused without modification. This motivates the second Lagrangian, introduced in Section~\ref{sec:lag2}, where the symmetry is restored by force. The price to pay is additional Lagrange multipliers, which now live in the tangent spaces of the variational manifold at the approximate CASCI vectors. These are not necessary to solve for when the ground-state energy is the sole point of interest. In the full CASCI limit, the first and second Lagrangians are equivalent, but when restricted to the variational manifold, the Lagrangians effecively describe different approximations.

The third Lagrangian, introduced in Section~\ref{sec:lag3}, is equivalent to the second Lagrangian, but has the added benefit that it is on the form of an expectation value. It is therefore suitable for formulating an action integral, while it also gives additional insight into the second Lagrangan's structure. The action-integral formulations of the first and third Lagrangians are discussed in Section~\ref{sec:action}, including their equations of motion.

Section~\ref{sec:applications} discusses some possible applications of the variational formulations, including iterative improvement of DMRG-tailored CC. Section~\ref{sec:conclusion} sums up and presents some future perspectives.

\section{The CASCC method}
\label{sec:cascc}

We discuss CASCC in sufficiently general terms as to be applicable to molecular systems, nuclear structure calculations, and even number-conserving boson systems with only small modifications. As usual in CC theory papers, we need to specify our notation and index conventions, which we attempt to keep at an economical level. We assume a standard setup with a computational Hilbert space $\mathcal{H}$  generated by a fixed set of single-particle functions (SPFs), i.e., $\mathcal{H}$ is the span of all possible linearly independent determinants (or permanents in the boson case) constructed with these functions. The computational Hilbert space is split into a model space $\mathcal{H}_0$  and its orthogonal complement $\mathcal{H}_\text{ext}$, the external space, i.e., $\mathcal{H} = \mathcal{H}_0 \oplus \mathcal{H}_\text{ext}$. The model space is assumed to be a standard complete-active space (CAS), and for simplicity, we assume that no SPFs are inactive. The determinantal basis for $\mathcal{H}$ is denoted $\{\ket{\phi_\rho}\}$, while the determinantal bases for $\mathcal{H}_0$ and $\mathcal{H}_\text{ext}$ are denoted $\{\ket{\phi_\alpha}\}$ and $\{\ket{\phi_\mu}\}$, respectively. Thus, we will use the convention that $\alpha,\beta,\ldots$ denote model space detereminant indices, $\mu,\nu,\ldots$ determine external-space indices, and $\rho,\sigma,\ldots$ are of either type.

In the fermion case, a set of $N$ CAS functions constitute the formal reference, while in the boson case one single  $\varphi_0\subset \{\varphi_p\}$ with occupation number $N$ determines the formal reference, in both cases denoted $\ket{\phi_0}$. We denote excitation operators relative to $\ket{\phi_0}$ by $X_\alpha$ (excitatons within the model space), and $X_\mu$ (external excitations, i.e., excitations with at least one non-CAS label). In general, we allow $\alpha=0$ in a model-space cluster operator, for which $X_\alpha=I$, the identity operator. A general excitation of either type is denoted $X_\rho$. The algebra of general cluster operators $T = \sum_\rho \tau_\rho X_\rho$ is denoted $\mathcal{T}$, and the subalgebra of model-space excitation operators $T_0 = \sum_\alpha \tau_\alpha X_\alpha$ is denoted $\mathcal{T}_0$. The set of external cluster operators $T = \sum_{\mu} \tau_\mu X_\mu$ is denoted $\mathcal{T}_\text{ext}$, and as linear spaces $\mathcal{T} = \mathcal{T}_0 \oplus \mathcal{T}_\text{ext}$.

We have  the resolution of identity $I = P + Q$ with
\[ P = \sum_{\alpha} \ket{\phi_\alpha}\bra{\phi_\alpha}, \quad Q = \sum_{\mu} \ket{\phi_\mu}\bra{\phi_\mu}. \]
Any $\ket{\psi}\in\mathcal{H}$ such that $\braket{\phi_0|\psi}\neq 0$ can now be written uniquely as $\ket{\psi} = \ket{\psi_0} + \ket{\psi_\text{ext}} =  (S_0 + S) \ket{\phi_0}$, with $S_0 \in \mathcal{T}_0$, and $S\in\mathcal{T}_\text{ext}$. Since any particle-number conserving cluster operator is nilpotent, there is a unique $T_0 + T \in \mathcal{T}$ such that $\exp(T_0 + T) = \exp(T)\exp(T_0) = S_0 + S$. Let $C = \exp(T_0) \in \mathcal{T}_0$ (since $\mathcal{T}_0$ is a subalgebra). It follows that
\begin{equation}
    \ket{\psi} = e^T \ket{\psi_0} = e^T C \ket{\phi_0}, \label{eq:cascc-ket}
\end{equation}
a unique decomposition and the basic ansatz for CASCC, and at the current stage it is equivalent to untruncated standard single-reference CC theory. 

Consider now the (partially) similarity transformed Schrödinger equation,
\begin{equation}
    e^{-T}He^T \ket{\psi_0} = E \ket{\psi_0},
    \label{eq:se}
\end{equation}
where $H$ is the system Hamiltonian. Defining $\bar{H}_\text{CAS} = P\bar{H}P = Pe^{-T}He^TP = PHe^TP$, and projecting onto CAS  and external determinants we obtain a non-Hermitian CASCI eigenvalue problem and external CC amplitude equations, respectively,
\begin{subequations}\label{eq:cascc-working-eq}
\begin{align}
    \bar{H}_\text{CAS}  \ket{\psi_0} &= E \ket{\psi_0}, \label{eq:cas-evp}\\
    \braket{\phi_\mu|\bar{H}|\psi_0} &= 0. \label{eq:ext-amp}
\end{align}
\end{subequations}
Equations~\eqref{eq:cas-evp} and \eqref{eq:ext-amp} are the linked CASCC working equations and fulfilled if and only if the exact Schrödinger equation $H\ket{\psi} = E \ket{\psi}$ is satisfied, as at this point, no cluster operator truncations have been introduced.  We remark, that some implementations of semilinear CASCC employs the equivalent unlinked formulation where the Schrödinger equation is not similarity transformed, and where Eq.~\eqref{eq:ext-amp} instead becomes $\braket{\phi_\mu | (H-E) \exp(T)|\psi_0}=0$.

The standard CASCC approach to approximate treatments is to truncate the external amplitude spaces (for both $T$ and $\Lambda$) to a discrete space $\mathcal{T}_d \subset \mathcal{T}_\text{ext}$ that includes, say, singles, doubles, and the first-order interaction space~\cite{Lyakh2012}. Furthermore, $T_0$ (and $\Lambda_0$) are truncated such that it includes a selected set of higher-order excitations that together allow spanning a sufficiently flexible model vector.  However, if the semilinear ansatz is considered, the CAS cluster operators must be untruncated. If not, there will be nonlinear constraints for the linear CAS amplitudes. (It is then typical to do a full similarity transform of $H$, replacing Eqs.~\eqref{eq:cascc-working-eq} with the fully linked SRCC equations $\braket{\phi_\rho|e^{-T-T_0}He^{T+T_0}|\phi_0} = E\delta_{\rho,0}$.) Thus, we assume in the following that the CASCI vectors are kept untruncated. As in standard SRCC, for each amplitude $t_\mu$ included in the approximate cluster operators $T\in \mathcal{T}_d$, the corresponding projective equation is included, constituting exactly one equation per unknown amplitude.

\section{The first Lagrangian}
\label{sec:lag1}

We observe that for any $\bra{\tilde{\psi}_0} \in \mathcal{H}_0^\dag$, (equivalently, for any $\tilde{C} \in \mathcal{T}_0^\dag$, with $\bra{\tilde{\psi}_0} = \bra{\phi_0}\tilde{C}$), such that $\braket{\tilde{\psi}_0|\psi_0}\neq 0$, we have
\begin{equation} 
E = \mathcal{E}_{\bar{H}_\text{CAS}}(\tilde{\psi}_0,\psi_0) \equiv \frac{\braket{\tilde{\psi}_0|\bar{H}_\text{CAS}|\psi_0}}{\braket{\tilde{\psi}_0|\psi_0}} \label{eq:energy}
\end{equation}
if Eq.~\eqref{eq:cas-evp} is fulfilled.
Indeed, it is readily verified that the bivariate Rayleigh quotient $\mathcal{E}_{\bar{H}_\text{CAS}}$ is stationary with respect to variations in $\bra{\tilde{\psi}_0}$ if and only if Eq.~\eqref{eq:cas-evp} is satisfied, with the additional constraint that $\braket{\tilde{\psi}_0|\psi_0} \neq 0$. The amplitude equations~\eqref{eq:ext-amp} are then additional constraints, leading to the introduction of a Lagrange multiplier cluster operator $\Lambda = \sum_\mu \lambda_\mu X_\mu^\dag \in \mathcal{T}_\text{ext}^\dag$ ($\in\mathcal{T}_d^\dag$ in the truncated case) and a Lagrangian
\begin{equation}
    L \equiv \braket{\tilde{\psi}_0|\psi_0}^{-1} (\bra{\tilde{\psi}_0} + \bra{\phi_0}\Lambda)\bar{H} \ket{\psi_0}. \label{eq:lag1}
\end{equation}
This is the first CASCC Lagrangian, providing a fully variational formulation whenever the model functions are  full CASCI vectors. Indeed, it is readily verified, that optimization of $L$ with respect to $\lambda_\mu$ and $\bra{\tilde{\psi}_0}$ reproduces the CASCC working equations~\eqref{eq:cas-evp} and \eqref{eq:ext-amp}, and that at the solution $L$ evaluates to the energy, regardless of the values of $\Lambda$ and $\bra{\tilde{\psi}_0}$. Differentiation of $L$ with respect to the variables $\ket{\psi_0}$ and $\tau_\mu$ gives an \emph{inhomogenous} CASCI eigenvalue problem and a linear system for $\Lambda$, respectively,
\begin{subequations}
\begin{align}
    \bra{\tilde{\psi}_0} \bar{H}_\text{CAS} +  \bra{\phi_0}\Lambda\bar{H} P&= E \bra{\tilde{\psi}_0} , \label{eq:cas-evp2}\\
    (\bra{\tilde{\psi}_0} + \bra{\phi_0} \Lambda)[\bar{H},X_\mu] \ket{\psi_0} &= 0. \label{eq:lambda-eq}
\end{align}
\end{subequations}
It is interesting to note that Eq.~\eqref{eq:lag1} is on the form  $L(\psi,\tilde{\psi}) = \braket{\tilde{\psi}|H|\psi}/\braket{\tilde{\psi}|\psi}$, with 
\begin{equation}
    \bra{\tilde{\psi}} = (\bra{\tilde{\psi}_0} + \bra{\phi_0} \Lambda) e^{-T}. \label{eq:cascc-bra}
\end{equation}
Since $e^{-T}$ is invertible, $\bra{\tilde{\psi}}\in\mathcal{H}^\dag$ is completely general. This shows that the Lagrangian $L$ can be interpreted in terms of Arponen's bivariational principle~\cite{Arponen1983}, and that the bra and ket will be exact left and right eigenvectors of $H$ in the full, untruncated limit. Moreover, the quantum mechanical state is assigned the density operator
\begin{equation}
    \rho = \braket{\tilde{\psi}|\psi}^{-1}\ket{\psi}\bra{\tilde{\psi}}.\label{eq:rho}
\end{equation}
When requiring the Hellmann--Feynman theorem to be valid, this implies that expectation values of observables are given by $\braket{\Omega} = \operatorname{Tr} (\rho\Omega) = \braket{\tilde{\psi}|\Omega|\psi}/\braket{\tilde{\psi}|\psi}$
whenever all variables $(T,\Lambda,\psi_0,\tilde{\psi}_0)$ are variationally determined as a critical point of $L$.

The equations~\eqref{eq:cas-evp2} and \eqref{eq:lambda-eq} are coupled, which in a practical calculation can be resolved as follows: Let $\bra{\psi'_0} \in \mathcal{H}_0^\dag$ be such that
\[ \bra{\psi'_0} \bar{H}_\text{CAS} = E \bra{\psi'_0}, \]
and we assume that the eigenvalue $E$ is isolated, and that $\braket{\psi_0'|\psi_0} = 1$. Introduce the projector $P_0 = \ket{\psi_0}\bra{\psi'_0}$, and observe that $\bar{H}_\text{CAS} = E P_0 + (1-P_0)\bar{H}_\text{CAS} (1-P_0)$. We assume now that $E$ is the leftmost eigenvalue of $\bar{H}_\text{CAS}$. Since $E$ is isolated, $\bar{H}_\bot \equiv (1-P_0)(\bar{H}_\text{CAS} - E)(1-P_0)$ has a right inverse as operator on the  subspace $\mathcal{H}_\bot^\dag = \mathcal{H}_0^\dag (1-P_0)$. We now decompose $\bra{\tilde{\psi}_0} = \bra{\psi'_0} + \bra{\chi}$ with $\bra{\chi} \in \mathcal{H}_\bot^\dag$, so that $\bra{\tilde{\psi}_0}$ is normalized as $\braket{\tilde{\psi}_0|\psi_0} = \braket{\psi'_0|\psi_0} = 1$. Inserting the decomposition into Eq.~\eqref{eq:cas-evp2} gives
\begin{equation}
    \bra{\chi} = -\bra{\phi_0}\Lambda \bar{H} (1-P_0) \bar{H}_\bot^{-1}.
\end{equation}
Equation~\eqref{eq:lambda-eq} becomes
\begin{equation}
    (\bra{\psi'_0} - \bra{\phi_0}\Lambda \bar{H} (1-P_0)\bar{H}_\bot^{-1} + \bra{\phi_0} \Lambda)[\bar{H},X_\mu]\ket{\psi_0} = 0,
\end{equation}
which is now  a linear system for $\Lambda$ in terms of known quantities.

\section{Variational approximations of the first Lagrangian}
\label{sec:variational}

Equation~\eqref{eq:lag1} describes a map $L : \mathcal{T}_d \times \mathcal{T}^\dag_d \times \mathcal{H}_0 \times \mathcal{H}_0^\dag \to \mathbb{C}$, whose critical points are equivalent to solutions of the left and right Schrödinger equations when $\mathcal{T}_d = \mathcal{T}_\text{ext}$ is not truncated. The CASCI vectors are kept untruncated, which incurs an exponential scaling of computational cost with respect to the system size, also for truncated cluster operators. We therefore introduce a differentiable manifold $\mathcal{M}\subset \mathcal{H}_0$ in which we wish to approximate $\ket{\psi_0} \in \mathcal{M}$ and $\bra{\tilde{\psi}_0} \in \mathcal{M}^\dag$ (where $\bra{\omega} \in \mathcal{M}^\dag$ if and only if $\ket{\omega} \in \mathcal{M}$), leading to a projected Lagrangian $L : \mathcal{T}_d \times \mathcal{T}_d^\dag \times \mathcal{M} \times \mathcal{M}^\dag \to \mathbb{C}$. Relevant examples to keep in mind are the Hartree--Fock manifold, i.e., the set of Slater determinants defined by $N$ linearly independent SPFs, considered here for its simplicity, and the manifold of matrix-product states (MPSs) of fixed bond dimension~\cite{Holtz2011,Bachmayr2016}.  We will henceforth let $\mathcal{M}$ be understood from context, and in particular the original exact case $\mathcal{M} = \mathcal{H}_0$ is a special case.

Near any $\ket{\psi_0}\in\mathcal{M}$ we can choose local coordinates $z \in V \equiv \mathbb{C}^n$, where $n$ is the dimension of $\mathcal{M}$. That is, there is a smooth map $\chi : V \to \mathcal{M}$ with a smooth inverse, with $\ket{\psi_0} = \ket{\chi(z_0)}$ for some $z_0\in V$. (Strictly speaking, the map $\chi$ may depend on the neighborhood of $\psi_0$ considered, as there may not be a global coordinatization of $\mathcal{M}$. However, this has no bearing on the results.) The tangent space $T_{\psi_0}\mathcal{M}$ is the $n$-dimensional subspace of $\mathcal{H}_0$ spanned by the tangent vectors $\ket{t_k} \equiv \ket{\partial_k\chi(z_0)}$, with $\partial_k \equiv \partial/\partial_{z_k}$. Without loss of generality, we may assume that these tangent vectors form an orthonormal system. Thus, the orthogonal projector onto $T_{\psi_0}\mathcal{M}$ is $P_{\psi_0} \equiv \sum_k \ket{t_k}\bra{t_k}$. It follows from orthonormality that $\ket{\partial_l t_k} = \ket{\partial_l\partial_k\chi(z_0)} = (I-P_{\psi_0}) \ket{\partial_lt_k}$. Similarly, we define local coordinates $\tilde{z}\in V$ such that $\bra{\tilde{\psi}_0} = \bra{\chi(\tilde{z}_0)}\in\mathcal{M}^\dag$, with corresponding tangent vectors $\bra{\tilde{t}_k}\equiv\bra{\partial_k\chi(\tilde{z}_0)}$ and projector $P_{\tilde{\psi}_0} = \sum_k \ket{\tilde{t}_k}\bra{\tilde{t}_k}$. We will assume without much loss of generality that $\mathcal{M}$ \emph{contains rays}, i.e., $\ket{\psi_0} \in \mathcal{M}$ implies $c\ket{\psi_0}\in\mathcal{M}$ for all scalars $c$. This again implies that $\ket{\psi_0} \in T_{\psi_0}\mathcal{M}$. For tangent vectors we will use the notation $\ket{v} = \sum_k v_k \ket{t_k}$. However, since $\ket{v}$ depends on its attachment point, a more proper notation would be $\ket{v;\psi_0}$ or similar. We will let it be clear from context whether a ket is in $\mathcal{M}$ or in $T_{\psi_0}\mathcal{M}$. 

We will shortly consider the stationary conditions of $L_{\mathcal{M}}$, for which we need to introduce arbitrary variations of of the model functions. We first take a detour, however, and consider the bivariational approximation of the left and right eigenvectors of an \emph{arbitrary} non-Hermitian operator $K$ by vectors in $\ket{\psi_0} \in \mathcal{M}$ and $\bra{\psi_0'} \in \mathcal{M}^\dag$. The different notation for the bra compared to the above is deliberate. We introduce  the \emph{bivariate Rayleigh quotient}
\begin{equation}
    \mathcal{E}_K(\psi_0,\psi_0') \equiv \frac{\braket{\psi_0'|K|\psi_0}}{\braket{{\psi}_0'|\psi_0}}.
\end{equation}
It is straightforward to compute its infinitesimal variation as
\begin{multline}
    \braket{{\psi}_0'|\psi_0}\delta\mathcal{E}_K(\psi_0,{\psi}_0') = \bra{\delta{\psi}_0'}(K - E)\ket{\psi_0} \\ + \bra{\psi_0'}(K-E) \ket{\delta\psi_0},
\end{multline}
where $E = \mathcal{E}_K(\psi_0,{\psi}_0')$. In the exact case, where $\mathcal{M}$ is the full linear space, $\mathcal{E}_K$ is clearly stationary if and only if $K\ket{\psi_0}=E\ket{\psi_0}$, $\bra{\psi_0'} K = E \bra{\psi_0'}$, and $\braket{\psi_0'|\psi_0} \neq 0$, which is an expression of the bivariational principle~\cite{Arponen1983,Lowdin1983}, i.e., a variational formulation of the simultaneous solution of the left and right eigenvalue problem for $K$. For the approximation on the manifold, $\delta\mathcal{E}_K=0$ for all possible variations in the model functions if and only if
$\braket{\psi_0'|\psi_0} \neq 0$ and \begin{subequations}\label{eq:nhevp}
\begin{align}
 P_{{\psi}_0'} (K-E) \ket{\psi_0} &=
 0,
\\
\bra{{\psi}_0'} (K-E) P_{\psi_0} &= 0 .
  \end{align}
\end{subequations}
This is a coupled set of nonlinear equations for the bra and ket model functions. In the case of the Hartree--Fock manifold, Eq.~\eqref{eq:nhevp} becomes the non-Hermitian Hartree-Fock method of Froelich and Löwdin~\cite{Froelich1983}. In the case of an MPS, one algorithm to solve Eq.~\eqref{eq:nhevp} is the \emph{alternating linear scheme} (ALS)~\cite{Bachmayr2016,Holtz2012} suitably generalized to independent bra and kets, i.e., to non-Hermitian problems. A similar generalization can be done for the \emph{modified ALS (MALS)}, which, for the non-Hermitian eigenvalue problem, collapses to the non-Hermitian density-matrix renormalization group method (DMRG)~\cite{Chan2005}. The DMRG algorithm, however, modifies the bond ranks of the MPS during its iterations, strictly speaking leaving the differentiable manifold world. This is a technical issue we will presently ignore. We further remark, that in typical situations, the tangent space projectors are not used on explicit form. Many manifolds $\mathcal{M}$ of interest are also parameterized 
locally with coordinates having redundancies, a statement true for both the Hartree--Fock and MPS examples~\cite{Lubich2013,Bachmayr2016}. This is resolved by introducing the mathematical structure of a \emph{principal bundle}, and usually leads to straight-forward formulations of tangent-space equations~\cite{Lubich2008,Lubich2015}.

The stationary conditions $\delta L=0$ are now straightforwardly computed as
\begin{subequations}
\begin{align}
    P_{\tilde{\psi}_0}(\bar{H}_\text{CAS}-E)\ket{\psi_0} &= 0  \label{eq:varcasevp} \\
    \bra{\tilde{\psi}_0}(\bar{H}_\text{CAS}-E) P_{\psi_0} &= -\bra{\phi}\Lambda\bar{H} P_{{\psi}_0}.\label{eq:inh}\\
    \braket{\phi_\mu|\bar{H}|\psi_0} &= 0 \label{eq:extamp2} \\
    (\bra{\tilde{\psi}_0} + \bra{\phi_0}\Lambda)[\bar{H},X_\mu]\ket{\psi_0} & = 0. \label{eq:lambdaeq2}
\end{align}\label{eq:cascc-crit}
\end{subequations}
Again, we note that the nonlinearities lead to couplings, such that in contrast to the standard CASCC equation~\eqref{eq:cas-evp}, the ket model vector cannot be solved independently of the bra model vector. Moreover any algorithm developed for the non-Hermitian eigenvalue problem, i.e., numerical solutions of Eq.~\eqref{eq:nhevp} such as the non-Hermitian DMRG algorithm, are not reusable for Eq.~\eqref{eq:cascc-crit} due to the inhomogeneity in Eq.~\eqref{eq:inh}. This motivates the second Lagrangian to be presented in the next section.

\section{The second Lagrangian}
\label{sec:lag2}

The second Lagrangian is based on the observation that, in the full CASCI version of CASCC, the bra model vector $\bra{\tilde{\psi}_0}$ can be chosen arbitrarily and still produce the right energy by projection, see Eq.~\eqref{eq:energy}. We therefore devise a scheme that forces it to be the variational left-eigenvector of $\bar{H}_\text{CAS}$, i.e., we force \emph{both} Eqs.~\eqref{eq:nhevp} to hold with $K = \bar{H}_\text{CAS}$. Thus, we obtain additional constraints with Lagrange multipliers $\ket{u} \in T_{\psi_0}\mathcal{M}$ and $\bra{\tilde{u}} \in T_{\tilde{\psi}_0}\mathcal{M}^\dag$, producing the Lagrangian
 \begin{equation}
     \tilde{L} \equiv L + \partial_{\tilde{\psi}_0}\mathcal{E}(\psi_0,\tilde{\psi}_0;\tilde{u}) + \partial_{\psi_0}\mathcal{E}(\psi_0,\tilde{\psi}_0;u),
     \label{eq:lag2}
 \end{equation}
with $\mathcal{E} = \mathcal{E}_{\bar{H}_{\text{CAS}}}$.
The last two terms are directional derivatives given by
\begin{equation}
    \partial_{\tilde{\psi}_0}\mathcal{E}(\psi_0,\tilde{\psi}_0;\tilde{u}) = \braket{\tilde{\psi}_0|\psi_0}^{-1}\bra{\tilde{u}}(\bar{H}_{\text{CAS}} - \mathcal{E}(\psi_0,\tilde{\psi}_0))\ket{\psi_0}, \label{eq:constr1}
\end{equation}
and
\begin{equation}
    \partial_{\psi_0}\mathcal{E}(\psi_0,\tilde{\psi}_0;u) = \braket{\tilde{\psi}_0|\psi_0}^{-1}\bra{\tilde{\psi}_0}(\bar{H}_{\text{CAS}} - \mathcal{E}(\psi_0,\tilde{\psi}_0))\ket{u}, \label{eq:constr2}
\end{equation}
respectively. In terms of the local coordinates, $\ket{u} = \sum_{k} \ket{t_k}\braket{t_k|u} = \sum_k \ket{t_k} u_k$ and $\bra{\tilde{u}} = \sum_k \tilde{u}_k \bra{\tilde{t}_k}$, with $u,\tilde{u}\in V$ being the $2n$ complex multipliers and the additional degrees of freedom introduced.

Differentiation with respect to the new Lagrange multipiers $\tilde{u}$ reproduces Eq.~\eqref{eq:varcasevp}, and differentiation with respect to $u$ produces the equation
\begin{equation}
\bra{\tilde{\psi}_0} (\bar{H}_\text{CAS} - E) P_{\psi_0} = 0, \tag{\ref{eq:inh}'} \label{eq:varcasevp2}
\end{equation}
replacing~Eq.~\eqref{eq:inh}. The amplitude equations~\eqref{eq:extamp2} are still obtained by differentiation with respect to $\lambda_\mu$. Thus, a non-Hermitian left/right eigenvalue calculation is coupled to the CASCC external amplitude equations. Once solved, the ground-state energy is obtained straightforwardly.

The multipliers $\Lambda$, $u$, and $\tilde{u}$ are required if properties other than the ground-state energy are to be computed. The equations for the multipliers are obtained by by differentiation with respect to $T$, $\bra{\tilde{\psi}_0}$, and $\ket{\psi_0}$, resulting in coupled linear systems, detailed in coordinate-free form in the Appendix.

\section{The third Lagrangian}

The second Lagrangian $\tilde{L}$ solved the problem of the inhomogeneity of the left eigenvalue problem. However, the functional is not of the form of an expectation value. Some squinting at Eq.~\eqref{eq:lag2} allows us to come up with the following functional, depending on the CASCC variables in addition to multipliers $\ket{v}\in T_{\psi_0}\mathcal{M}$ and $\bra{\tilde{v}} \in T_{\tilde{\psi}_0}\mathcal{M}^\dag$,
\label{sec:lag3}
\begin{multline}
    \hat{L}
    \equiv  \frac{1}{2}\braket{\phi_0|\Lambda\bar{H}|\psi_0} + \frac{1}{2}\frac{\braket{\tilde{v}|\bar{H}_\text{CAS}|\psi_0}}{\braket{\tilde{v}|\psi_0}} + \frac{1}{2} \frac{\braket{\tilde{\psi}_0|\bar{H}_\text{CAS}|v}}{\braket{\tilde{\psi}_0|v}}  .
\end{multline}
It is straightforward to show, that differentiation with respect to the tangent vectors reproduce Eqs.~\eqref{eq:varcasevp} and \eqref{eq:varcasevp2} by means of the bivariational principle. Differentiation with respect to $\lambda_\mu$ also reproduces Eq.~\eqref{eq:extamp2}. The factors $1/2$ now ensure that the Lagrangian evaluates to the energy at the critical point, and that we have $\hat{L} = \operatorname{Tr}(H\rho)$ with the rank-2 non-Hermitian density operator
$\rho = (\rho_1+\rho_2)/2$, where
\begin{equation}
    \rho_1 = \braket{\tilde{\psi_0}|v}^{-1} e^T \ket{v} \bra{\tilde{\psi}_0} e^{-T}\label{eq:dens1}
\end{equation}
and
\begin{equation}
    \rho_2 = \braket{\tilde{v}|\psi_0}^{-1} e^T \ket{\psi_0} (\bra{\tilde{v}} + \braket{\tilde{v}|\psi_0}\bra{\phi_0}\Lambda) e^{-T}. \label{eq:dens2}
\end{equation}
It is readily verified, that $\operatorname{Tr}(\rho_i) = 1$ and that $\rho_i^2=\rho_i$.

Thus, the Lagragians $\tilde{L}$ and $\hat{L}$ are equivalent, in the sense that the working equations are identical for $\ket{\psi_0}$, $\bra{\tilde{\psi}_0}$, and $T$, the values of which are sufficient to compute the energy using either Lagrangian. This also implies that the Hellmann--Feynman approach to expectation values and other properties yield identical results, even if the Lagrange multipliers have different values for the two Lagrangians.

In the Appendix, the Lagrange multiplier equations are derived, and they are closely related to those of the second Lagrangian $\tilde{L}$, with basically the same computational footprint and details. 

\section{Time-dependent formalism}
\label{sec:action}

Real-time propagation of coupled-cluster theory is gaining interest in the research community. It offers a computational alternative to response theory and excited states, and there is a demand for accurate first-principles simulations of molecular systems subject to intense and ultrashort laser pulses, see Refs.~\cite{Li2020,Ofstad2023} for reviews. In this section, we provide an overview of the explicitly time-dependent generalizations of the variational formulations. A detailed exposition including response theory based on this formalism is relegated to future work. 

\subsection{Action functional for the first Lagrangian}

The time-dependent Schrödinger equation and its dual are the critical point conditions of the bivariate action functional~\cite{Chernoff1974,Arponen1983}
\begin{equation}
    S \equiv \int_{t_0}^{t_1} \braket{\tilde{\psi}(t)| i \partial_t - H | \psi(t)} \, dt, \label{eq:action}
\end{equation}
depending on the history of the system state for time $t\in [t_0,t_1]$. As usual, any infinitesimal variation is assumed vanishing at the endpoints of the interval. Thus, $\delta S=0$ with respect to variations in $\bra{\tilde{\psi}}$ and $\ket{\psi}$ if and only if, respectively,
\begin{equation}
    H\ket{\psi(t)} = i \partial_t \ket{\psi(t)}, \,\text{and}\,
\bra{\tilde{\psi}(t)} H = -i \partial_t \bra{\tilde{\psi}(t)},
\end{equation}
in addition to boundary conditions. The functional $S$ generalizes the expectation-value functional to the time domain by treating the energy as generator for time evolution. Interestingly, the equations are \emph{canonical}, i.e., on the form of classical Hamiltonian mechanics,
\begin{equation}
    i \partial_t \ket{\psi} = \partial_{{\tilde{\psi}}} \mathcal{K}, \quad -i \partial_t \bra{\tilde{\psi}} = \partial_{{\psi}} \mathcal{K},
\end{equation}
with the Hamiltonian function $\mathcal{K} = \braket{\tilde{\psi}|H|\psi}$. We remark, that the overlap $\braket{\tilde{\psi}|\psi}$ is preserved during time evolution.

The functional~\eqref{eq:action} is appropriate also for approximate states that contain rays, which is true for the CASCC bra and ket: For arbitrary complex scalars $c$ and $\tilde{c}$, $\ket{\psi}\to c\ket{\psi}$ and $\bra{\tilde{\psi}} \to \tilde{c}\bra{\tilde{\psi}}$ under the transformations $\ket{\psi_0}\to c\ket{\psi_0}$ and $(\Lambda,\bra{\tilde{\psi}_0}) \to (\tilde{c}\Lambda,\tilde{c}\bra{\tilde{\psi}_0})$. Thus, $\rho$ in Eq.~\eqref{eq:rho} is invariant. We can therefore assume that $\operatorname{Tr}\rho(t_0)=1$, which will be preserved during evolution. For approximate methods that do not contain rays, the functional must be made explicitly invariant with respect to time-local phase and normalization by dividing the integrand by $\braket{\tilde{\psi}(t)|\psi(t)}$, as detailed for traditional variational methods in Ref.~\cite{Kramer1981}.

Inserting the CASCC bra and ket vectors from Eqs.~\eqref{eq:cascc-ket} and \eqref{eq:cascc-bra}, we obtain the functional
\begin{equation}
    S = \int_0^T i \braket{\phi_0|\Lambda\dot{T}|\psi_0} + i \braket{\tilde{\psi}_0|\dot{\psi}_0} - \mathcal{K}(\tau,\lambda,\tilde{\psi}_0,\psi_0) \, dt,
\end{equation}
where the dot denotes a time derivative, and where
\begin{equation}
    \mathcal{K}(T,\Lambda,\psi_0,\tilde{\psi}_0) = (\bra{\phi_0}\Lambda + \bra{\tilde{\psi}_0})\bar{H}\ket{\psi_0},
\end{equation}
is the energy expressed in the CASCC variables, i.e., the Lagrangian~\eqref{eq:lag1} without the denominator. 
We are assuming that at all times $t$, the model vectors are $\ket{\psi_0(t)} \in \mathcal{M}$, and $\bra{\tilde{\psi}_0(t)} \in \mathcal{M}^\dag$. Requiring $S$ to be stationary with respect to all independent variables now reveals equations of motion of the form
\begin{subequations}\label{eq:lag1-tdse}
\begin{align}
   i P_{\tilde{\psi}_0} \partial_t \ket{\psi_0} &= P_{\tilde{\psi}_0} \bar{H} \ket{\psi_0} \label{eq:right-tdse}\\
 i X(\psi_0) \dot{\tau} &= F(\tau,\psi_0), \label{eq:amp-tdse} \\
  - i (\partial_t \bra{\tilde{\psi}_0}) P_{\psi_0} &=
  - i \bra{\phi_0} \Lambda \dot{T} P_{\psi_0} + (\bra{\tilde{\psi}_0} + \bra{\phi_0}\Lambda)\bar{H}P_{\psi_0}.
  \label{eq:left-tdse-inh} \\  - i X(\psi_0)^T \dot{\lambda} &=  G_0(\tau,\lambda,\psi_0,\tilde{\psi}_0) + G(\tau,\lambda,\psi_0,\tilde{\psi}_0), \label{eq:lambda-tdse} 
 \end{align}
\end{subequations}
the derivation of which is very similar to the stationary conditions for the Lagrangian $L$. The vector-valued functions $F$ and $G$ are the Lagrangian derivatives with respect to $\lambda$ and $\tau$, respectively, under the condition that $\braket{\tilde{\psi}|\psi}=1$ (which is preserved during evolution), i.e.,
\begin{align} F(\tau,\psi_0)_\mu &= \braket{\phi_\mu|\bar{H}|\psi_0},\\
 G(\tau,\lambda,\psi_0,\tilde{\psi}_0)_\mu &= (\bra{\tilde{\psi}_0} + \bra{\phi_0}\Lambda) [\bar{H},X_\mu]\ket{\psi_0}.
\end{align}
The vector-valued function $G_0$ is given by
\begin{equation}
    G_0(\tau,\lambda,\psi_0,\tilde{\psi}_0)_\mu =  \braket{\phi_0|\Lambda X_\mu J \bar{H}|\psi_0}.
\end{equation}
Here, we have introduced an oblique projection operator $J=J(\psi_0,\tilde{\psi}_0)$ discussed in the Appendix.
The occurence of $\dot{T}$ in Eq.~\eqref{eq:left-tdse-inh} can be eliminated by means of Eq.~\eqref{eq:amp-tdse}. In the case where $\mathcal{M} = \mathcal{H}_0$, i.e., the full CASCI model spaces, $P_{\psi_0} = P_{\tilde{\psi}_0} = P_\text{CAS}$.
The equations of motion should be compared with the CASCC working equations. In particular, we note the non-Hermitian right  time-dependent Schrödinger equation~\eqref{eq:right-tdse} and an inhomogenous left equation~\eqref{eq:left-tdse-inh}. We may ``invert'' the projector $P_{\tilde{\psi}_0}$ by means of $J$ (see the Appendix), so that Eq.~\eqref{eq:right-tdse} becomes
\begin{equation}
    i\partial_t \ket{\psi_0} = J \bar{H} \ket{\psi_0},\tag{\ref{eq:right-tdse}'}
\end{equation}
and Eq.~\eqref{eq:left-tdse-inh} becomes
\begin{equation}
 - i \partial_t \bra{\tilde{\psi}_0} =
 - i \bra{\phi_0} \Lambda \dot{T} J + (\bra{\tilde{\psi}_0} + \bra{\phi_0}\Lambda)\bar{H} J.
    \tag{\ref{eq:left-tdse-inh}'}
\end{equation}
We also note, that with a critical point of $L$ as initial condition, e.g., the ground state, the solutions to Eqs.~\eqref{eq:lag1-tdse} are easily integrated to a stationary state given by $\ket{\psi(t)} = \exp(-iEt)\ket{\psi(0)}$ and $\bra{\tilde{\psi}(t)} = \exp(iEt)\bra{\tilde{\psi}(0)}$. Linear perturbation theory will yield response-theory type results~\cite{McWeeny1992}.

\subsection{Action functional for the third Lagrangian}

Turning to the third Lagrangian, the pure state is replaced by the density operator $\rho = (\rho_1+\rho_2)/2$, with $\rho_i = \ket{\psi_i}\bra{\tilde{\psi}_i}$ being pure-state operators, see Eqs.~\eqref{eq:dens1} and \eqref{eq:dens2}. For a general convex combination $\rho = \sum_k p_k \rho_k$ of this form, where $0\leq p_k \leq 1$ and $\sum_k p_k = 1$, we \emph{define} an action functional
\begin{equation}
    \hat{S} = \sum_k \int_0^T p_k \bra{\tilde{\psi}_k}i \partial_t - H \ket{\psi_k} \, dt.
\end{equation}
Equations of motion for $\rho_i$ are now obtained by forcing $\hat{S}$ to be critical. It is readily seen, that if $E = \operatorname{Tr}(\rho H)$ is critical, then $\rho(t)=\rho$ gives a critical point of $\hat{S}$, as expected. Moreover, if all the pure states are independent and allowed to vary freely in the full computational Hilbert space $\mathcal{H}$, then $\hat{S}$ is critical if and only if the left and right time-dependent Schrödinger equations for each pair $\ket{\psi_i}$ and $\bra{\tilde{\psi}_i}$ are satisfied. Finally, for approximate bra and ket wavefunctions whose normalization is not fixed, i.e., the manifold of $(\psi_i,\tilde{\psi}_i)$ contains rays, one can show that $\braket{\tilde{\psi}_i|\psi_i}$ is preserved during evolution, and in particular that $\operatorname{Tr}(\rho)$ is a constant of motion. In conclusion, the action functional $\hat{S}$ is a reasonable starting point for defining approximate time evolution.

For the  ansatz $\rho = (\rho_1+\rho_2)/2$ of the third Lagrangian, we may now assume $\braket{\tilde{\psi}_0|v}=\braket{\tilde{v}|\psi_0}=1$, which yields
\begin{equation}
    \hat{S} = \frac{1}{2}\int_0^T i\braket{\phi_0|\Lambda\dot{T}|\psi_0} + i \braket{\tilde{v}|\dot{\psi}_0} + i\braket{\tilde{\psi}_0|\dot{v}} + i\braket{\tilde{\psi}_0|\dot{P}_{\psi_0}|v} -2\hat{\mathcal{K}}, \label{eq:action-lag3}
\end{equation}
with
\begin{equation}
    \hat{\mathcal{K}} = \frac{1}{2}\left(\braket{\phi_0|\Lambda \bar{H}|\psi_0} + \braket{\tilde{v}|\bar{H}|\psi_0} + \braket{\tilde{\psi}_0|\bar{H}|v}\right)
\end{equation}
being obtained from the Lagrangian $\hat{L}$ with the assumed normalizations. 
Using integration by parts, we obtain the equivalent action
\begin{equation}
    \hat{S} = \frac{1}{2}\int_0^T i\braket{\phi_0|\Lambda\dot{T}|\psi_0} -
    i \braket{\dot{\tilde{v}}|{\psi}_0} - i\braket{\dot{\tilde{\psi}}_0|v} - i\braket{\tilde{v}|\dot{P}_{\tilde{\psi}_0}|\psi_0} -2\hat{\mathcal{K}}, \label{eq:action-lag3-prime} \tag{\ref{eq:action-lag3}'}
\end{equation}
indicating that the equations of motion of the bras and kets will exhibit a certain symmetry.

Using similar techniques as outlined in the Appendix, we can now derive equations of motion. Variation of the Lagrange multipliers $\Lambda$ and $\tilde{v}$ reproduces Eqs.~\eqref{eq:right-tdse} and \eqref{eq:amp-tdse}. Variation of the Lagrange multiplier $v$ gives the equation
\begin{subequations}
    \begin{align}
- i \bra{\dot{\tilde{\psi}}_0} P_{\psi_0} &= \bra{\tilde{\psi}_0} \bar{H} P_{\psi_0},\label{eq:left-tdse}
    \end{align}
which replaces Eq.~\eqref{eq:left-tdse-inh}, in a similar manner as for the time-independent case. Continuing, variation of $T$ gives the equation
\begin{align}
    -i A(c)^T \dot{\lambda} &= \frac{1}{2} G_0(\tau,\lambda,\psi_0,\tilde{\psi}_0) +  \hat{G}(\tau,\lambda,\psi_0,\tilde{\psi}_0,v,\tilde{v}) ,
\end{align}
\end{subequations}
which replaces Eq.~\eqref{eq:lambda-tdse}. Here,
\begin{equation}
    \hat{G}_\mu\partial_{\tau_\mu} \hat{\mathcal{K}}  = \frac{1}{2}\braket{\phi_0|\bar{H}_\mu|\psi_0} + \frac{1}{2}\braket{\tilde{v}|\bar{H}_\mu|\psi_0} + \frac{1}{2}\braket{\tilde{\psi}_0|\bar{H}_\mu|v},
\end{equation}
with $\bar{H}_\mu = [\bar{H},X_\mu]$, see the Appendix.

It remains to determine equations of motion for the Lagrange multipliers $v$ and $\tilde{v}$, obtained by varying $\bra{\tilde{\psi}_0}$ and $\ket{\psi_0}$. Unlike the time-independent case, the multipliers are necessary to compute any expectation value, i.e., any physical prediction. The equations of motion, whose derivation is outlined in the Appendix, read

\begin{subequations} \label{eq:multiplier-eom-lag3}
\begin{equation} \label{eq:multiplier-eom-lag3-a}
    \begin{split}
        -i \bra{\dot{\tilde{v}}} &=  \bra{\phi_0}\Lambda(\bar{H}-i\dot{T})J + \bra{\tilde{v}} \bar{H} J \\ &\quad -  \bra{\tilde{\psi}_0} \bar{H} (J P'_{\tilde{\psi}_0}(\tilde{v}) + [J,P'_{\psi_0}(v)] ) J     ,
    \end{split} 
\end{equation}
and
\begin{equation}
    \begin{split}
        i \ket{\dot{v}} &=  J \bar{H}\ket{v}  -  J ( P'_{\psi_0}(v) J + [P'_{\tilde{\psi}_0}(\tilde{v}),J]) \bar{H} \ket{\psi_0}.
    \end{split}
\end{equation}
\end{subequations}
The occurence of $\dot{T}$ in Eq.~\eqref{eq:multiplier-eom-lag3-a} can be eliminated using the amplitude equation of motion.

While these equations seem fairly complicated, they are expressed on coordinate-free form, and are expected to simplify considerably when concrete manifold parameterizations are employed. Moreover, if a manifold $\mathcal{M}$ with low dimension compared to the full model CAS, the overall cost is small compared to evolution of the amplitude equations.

\section{Possible applications}
\label{sec:applications}

While the Lagrangians derived in this article should be useful for standard CASCC calculations, especially when properties and explicit time evolution are considered, the generalization to CASCC manifold approximation can have several additional uses. We here outline only a few ideas to be explored in future research.

\subsection{Correction of tailored CC}
\label{sec:tcc}

The tailored CC method~\cite{Kinoshita2005} is an approach for external correction of a fixed variational model wavefunction. A variational model wavefunction $\ket{\psi_0} \in \mathcal{M} \subset \mathcal{H}_0$ is computed using, say CASCI as in Ref.~\cite{Kinoshita2005}, or the quantum chemical DMRG method, QC-DMRG~\cite{White1999,Legeza2003,Chan2004,Marti2010,Wouters2014}, and inserted into the CC amplitude equations for an external cluster operator $T \in \mathcal{T}_d$, producing the DMRG-TCC method~\cite{Veis2016,Faulstich2019,Faulstich2019b,Lang2020}. An important caveat is that only CAS singles and doubles are used in practice due to the cost of extracting triples from the DMRG wavefunction. Thus, one first solves
\begin{equation} \label{eq:tcc-cas}
     E_0 = \braket{\psi_{0,*}|H_\text{CAS}|\psi_{0,*}} = \min \left\{\braket{\psi|H_\text{CAS}|\psi} \mid \psi \in \mathcal{M} \right\},
\end{equation} 
or equivalently, 
\begin{equation}
    P_{\psi_{0,*}} (H_\text{CAS} - E_0) \ket{\psi_{0,*}} = 0,
\end{equation}
extracts an approximation $\ket{\psi_0} \approx \ket{\psi_{0,*}}$ due to practical considerations,
and subsequently solves Eq.~\eqref{eq:ext-amp} for the external $T\in\mathcal{T}_d$. Here, $H_\text{CAS} = PHP$. The total energy is then given by
\[ E_\text{TCC} = E_0 + E_\text{corr} = \mathcal{E}_{\bar{H}_{\text{CAS}}}(\psi_0,\psi_0),\]
see Eq.~\eqref{eq:energy}. In many cases, it is the fully exponential parameterization which is actually implemented, i.e., $\ket{\psi_0} = \exp[T_0]\ket{\phi_0} \approx \ket{\psi_{0,*}}$, with $T_0$ containing, say, only singles and doubles extracted from $\ket{\psi_{0,*}}$.

Clearly, the TCC method can be viewed as a first approximation to the self-consistent solution of Eq.~\eqref{eq:cascc-crit}, the stationary conditions of the first Lagrangian. Assume that $\ket{\psi_0} = \ket{\psi_{0,*}}$, i.e., no approximations to the TCC CAS reference is done when inserting into the external amplitude equations. In order to \emph{improve} upon the TCC calculation, then,  it is easy to see that the current estimate $(\ket{\psi_0},T)$ must be fed back into the critical point condition and iterated until convergence. Note, however, that subsequent iterations involve both a non-hermitian right eigenvalue problem and an inhomogenous left eigenvalue problem. 

Alternatively, one may resort to the second Lagrangian, where the inhomogenous left eigenvalue problem is replaced with a homogenous eigenvalue problem at the expense of additional Lagrange multipliers. However, \emph{if only the ground-state energy} is required, these multipliers are not needed. A two-sided non-Hermitian eigenvalue problem can then be computed variationally, selfconsistently with the $T$-amplitude equations, and the energy computed via
\[ E_\text{TCC} =  \mathcal{E}_{\bar{H}_{\text{CAS}}}(\tilde{\psi}_0,\psi_0), \]
generalizing the previous formula.
This approach allows existing implementations of non-Hermitian DMRG algorithms to be reused for corrected DMRG-TCC calculations. It must be stressed, however, that if additional properties need to be calculated, the Lagrange multipliers $\Lambda$, $u$ and $\tilde{u}$ need to be computed. 

The ideal approach, however, would be to use the first Lagrangian and adapt the DMRG algorithm to optimization of $L$. The microiterations in the DMRG sweeps would be smaller-dimension versions of Eqs.~\eqref{eq:varcasevp} and \eqref{eq:inh}, and would need novel computer implementations. We relegate a study of this algorithm for future study.

If some approximation to $\ket{\psi_{0,*}}$ is done, such as for current DMRG-TCC implementations, further considerations must be made. The semilinear CASCC Lagrangians still provide the exact result, and it is likely that correction schemes still can be found based on iterations or, e.g., perturbation theory.

\subsection{Attosecond dymanics}

With the advent of technology producing ultra-short laser pulses, it has been possible to experimentally probe the laws of quantum mechanics at a time-scale resolving the motion of electrons in complex molecules~\cite{Kaneshima2018}. In particular, even for an initial condition of single-reference nature, time evolution may produce strong multireference character. Moreover, description of unbound systems mandate the use of global resolution of single-particle basis functions in the form of grids, discrete-variable representations, or similar.  Currently, the state-of-the-art methods for such descriptions with time-dependent single-particle bases functions are time-dependent multiconfigurational Hartree--Fock-type methods (MCTDHF)~\cite{Zanghellini2003,Alon2007,Miyagi2013,Sato2015,Hochstuhl2014}, or single-reference coupled-cluster-based approaches~\cite{Kvaal2012,Sato2018,Pathak2020,Pathak2020a}. The CASCC Lagrangians may be excellent starting points for development of low-scaling methods applicable to larger molecules and with higher-quality static correlation description than before.

Let us sketch the procedure for the derivation of the orbital-adaptive time-dependent version of the first Lagrangian, for simplicity restricting our attention to the case $\mathcal{M}=\mathcal{H}_0$, i.e., we use the full, linear CASCI expansion for model vectors. The outline closely follows that of Ref.~\cite{Kvaal2012} in the general principles.

In an orbital-adaptive bivariational method such as we are going to describe for CASCC, the bra and ket are constructed from biorthogonal but otherwise independent sets $\{\tilde{\varphi}_x\}$ and $\{\varphi_x\}$ of SPFs, respectively, i.e., $\braket{\tilde{\varphi}_x|\varphi_y} \equiv \delta_{xy}$, and correspondingly the creation and annihilation operators associated with these functions satisfy $\{\tilde{c}_x,c^\dag_y\}=\delta_{xy}$. For bosons, the anticommutator is replaced by a commutator. 
We stress, that the individual SPFs are typically resolved in a much larger computational basis, e.g., some kind of grid or discrete-variable representation. This is \emph{mandatory} feature of real-time propagation methods whenever an unbound system of electrons are described, as a given finite and rather small basis set $\{\varphi_x\}$ and $\tilde{\varphi}_x$ is always localized,. Thus, the operator $P = \sum_x \ket{\varphi_x}\bra{\tilde{\varphi}_x}$ projects onto the current single-particle basis span, and $Q = 1 - P$ projects onto the remaining degrees of freedom. Typically, the range of $Q$ has \emph{much} higher dimension than the relatively compact single-particle space, and is not localized.

The SPF spaces are divided into occupied and unoccupied CAS functions $\{\varphi_i\}$ and $\{\varphi_a\}$, respectively, and external SPFs $\{\varphi_\alpha\}$, i.e., $\{\varphi_x\} = \{\varphi_i\}\cup\{\varphi_a\}\cup\{\varphi_x\}$, and similarly $\{\tilde\varphi_x\} = \{\tilde\varphi_i\}\cup\{\tilde\varphi_a\}\cup\{\tilde\varphi_x\}$.  An infinitesimal change in the SPFs that respect the biorthogonality constraint now reads~\cite{Kvaal2012}
\begin{align}
    \delta \ket{\varphi_y} &= \sum_x U_{yx} \ket{\varphi_x} + \ket{\chi_y},   \\
    \delta \bra{\tilde{\varphi}_x} &= -\sum_y \bra{\varphi_y} {U}_{yx}  + \bra{\tilde{\chi}_x},
\end{align}
where $U$ is a matrix, and 
where $\ket{\chi_y}$ and $\bra{\tilde{\chi}_x}$ are arbitrary SPFs such that $P\ket{\chi_y}=0$ and $\bra{\tilde{\chi}_x}P = 0$. Thus, the infinitesimal variations can be decomposed into a part along the basis itself, and an ``orthogonal'' part. This orthogonal part is responsible for describing ionization dynamics, for example, as it allows the SPFs to travel away from the local region in space described by the current value for the of SPF basis.

The CASCC state $\rho = \ket{\psi}\bra{\tilde{\psi}}$ is invariant under arbitrary transformations of the individual collections of occupied $\{\varphi_i\}$ CAS SPFs, the unoccupied $\{\varphi_a\}$ CAS SPFs, and the external $\{\varphi_\alpha\}$ SPFs. This means that the corresponding diagonal blocks in $U$  are redundant and can be set to arbitrary values, fixing gauge choices for the SPF variations. The remaining blocks and the functions $\bra{\tilde{\chi}}$ and $\ket{\chi}$ are independent variables of the variations, and when introduced into the action they will lead to equations of motion for $P\ket{\varphi_x}$,  $Q\ket{\varphi_x}$, $\bra{\tilde{\varphi}_x}P$, and $\bra{\tilde{\varphi}_x}Q$. 

The equations of motion are obtained as follows. First, we write $\ket{\psi} = e^T C\ket{\phi_0}$ where $C \in \mathcal{T}_0$ is a model-space cluster operator $C = \sum_\alpha c_\alpha X_\alpha$, and observe that the time derivative can be separated in terms of amplitude derivatives and SPF derivatives as
\[ \frac{d}{dt}\ket{\psi} = (D_{\text{amp}} + D_{\text{SPF}} )\ket{\psi}, \]
with $D_{\text{amp}}=\sum_\alpha \dot{c}_\alpha \frac{\partial}{\partial c_\alpha} + \sum_\mu \dot{\tau}_\mu \frac{\partial}{\partial \tau_\mu}$ and $D_{\text{SPF}} =  \sum_x \dot{c}^\dag_x \tilde{c}_x$.
When inserted into the action functional~\eqref{eq:action}, we get
\begin{equation}
    S = \int_{t_0}^{t_1} i \braket{\tilde{\psi}|\dot{\psi}_{\text{amp}}} - \braket{\tilde{\psi}|H - i D_{\text{SPF}}|\psi}\, dt,
\end{equation}
where $\ket{\dot{\psi}_{\text{amp}}} = D_{\text{amp}}\ket{\psi}$. We further mote note that $iD_\text{SPF}$, often called the Coriolis operator, acts as a one-body operator shift of the Hamiltonian, simply due to the time-dependence of the basis. Derivation of equations of motion for the amplitudes of $T$, $\Lambda$, and $\ket{\psi_0}$ and $\bra{\tilde{\psi}_0}$ now follows as before. For the derivation of the equations of motion for the SPFs, we rewrite the action as
\begin{equation}
    S = \int_{t_0}^{t_1} i \operatorname{Tr}({\rho^{(1)} \eta}) - \sum_{n=1,2}\operatorname{Tr}(\rho^{(n)} h^{(n)}) + i\braket{\tilde{\psi}| \dot{\psi}_{\text{amp}}}\, dt,
\end{equation}
where $\rho^{(n)}$ are reduced density matrices, being functions of the the amplitudes \emph{only} due to Wick's theorem, and where $\eta = [\eta_{xy}] = [\braket{\tilde{\varphi}_x|\dot{\varphi}_y}]$, $h^{(1)} = [\braket{\tilde{\varphi}_x|H^{(1)}|\varphi_{y}}]$ and $h^{(2)} = [\braket{\tilde{\varphi}_x\tilde{\varphi}_{x'}|H^{(2)}|\varphi_{y}\varphi_{y'}}]$, the integrals of the one-plus-two-body Hamiltonian $H = H^{(1)} + H^{(2)}$, and all these matrices are functions of the SPFs \emph{only}. Finally, $\braket{\tilde{\psi}|\dot{\psi}_{\text{amp}}}$ is a function of the amplitudes \emph{only} -- its variation is identically zero and it can be omitted from the action functional in this context. Variations of $\bra{\tilde{\varphi}_x}$ and $\ket{\varphi_y}$ on the form discussed above now straightforwardly produces (partially coupled) equations for $\ket{\dot{\varphi}_x}$ and $\bra{\dot{\varphi}_y}$, mean-field type equations formally similar to MCTDHF equations of motion.

The linear CAS model vector expansions may of course be replaced by, say, time-dependent matrix-product states. In that case, one has to carefully analyze the SPF invariance properties of the resulting state before deriving equations of motion.

\section{Conclusion}
\label{sec:conclusion}

We have introduced fully variational formulations of CASCC, in three different formulations: The standard CASCC method in its semilinear linked formulation, including a formulation where the CAS model functions are approximated on a smooth submanifold, and two equivalent yet different formulations, where the inhomogenous non-Hermitian left eigenvalue problem for the CAS problem is replaced by a homogenous eigenvalue problem. This has the \emph{benefit} that it partially decouples nonlinear working equations and allows reuse of already existing codes for the non-Hermitian two-sided eigenvalue problem, but the \emph{drawback} that it introduces additional variables, Lagrange multipliers living in tangent space of the CAS manifold.

The equations derived are fairly abstract, yet very general, and in actual applications, the CAS manifold, such as a matrix-product state, will have a quite complex definition and with a non-trivial description of the tangent space due to gauge degrees of freedom~\cite{Lubich2015}. Thus, the equations are expected to take on a different form in concrete realizations.

As for future research, several interesting projects are immediate. From the theory side, a fully variational framework can be used to derive response theory and excited states. Another interesting line of research is to investigate the extension of the present formalism from CASCC to the somewhat more general approach of subsystem-embedding subalgebras by Kowalski~\cite{Kowalski2018}.

On the practical side, the iterative correction of DMRG-TCC is an easy target, and can be approached by either the first or second/third Lagrangian, depending on whether one wants to reuse existing non-Hermitian DMRG implementations. One interesting approach is to generalize the DMRG algorithm to direct optimization of the first CASCC Lagrangian, by optimizing $L$ in sweeps, freezing all but uo to two blocks of the bra and ket MPSs in each microiteration. This approach would be a generalization of the alternating linear scheme (ALS) and the modified ALS algorithm~\cite{Holtz2012,Bachmayr2016} to bivariational problems.

In this article, no attempt at evaluating the computational cost of methods like CASCC with matrix-product states/DMRG has been attempted. Such considerations are complex and depend on implementation details of both CC amplitude equations and the DMRG algorithm for non-Hermitian operators. It is possible that implementations will be challenging, and that pragmatic considerations will force through  approximations not considered here. Moreover, even iterative corrections to DMRG-TCC using the Lagrangians may turn out to be prohibitively expensive. Even so, it is likely that some corrections can be extracted that are computationally affordable.

\acknowledgments

This work has received funding from the Research Council of Norway (RCN) under CoE Grant No 262695 (Hylleraas Centre for Quantum Molecular Sciences) and from ERC-STG-2014 under grant agreement No 639508 (BIVAQUM).


\appendix
\section{Working equations for the second Lagrangian}

A general variation in $\ket{\psi_0}=\ket{\chi(z_0)}\in\mathcal{M}$ can be written \begin{equation}
    \ket{\delta\psi_0} = \sum_k \ket{t_k}\delta z_k,
\end{equation}
and similarly
\begin{equation}
    \bra{\delta\tilde{\psi}_0} = \sum_k \delta\tilde{z}_k \bra{\tilde{t}_k}.
\end{equation}
General tangent vectors are $\ket{u} = \sum_k \ket{t_k}u_k$ and $\bra{\tilde{u}} = \sum_k \tilde{u}_k\bra{\tilde{t}_k}$.

We begin by computing the variation of the second Lagrangian $\tilde{L}$ in Eq.~\eqref{eq:lag2} with respect to $\bra{\tilde{\psi}_0}$. The variation  can be split into the variation of $L$ and of the directional derivative terms in Eqs.~\eqref{eq:constr1} and \eqref{eq:constr2}. The variation of $L$ with respect to $\bra{\tilde{\psi}_0}$ is
\begin{equation}
    \delta L = \braket{\tilde{\psi}_0|\psi_0}^{-1}\left(\bra{\delta\tilde{\psi}_0} \bar{H} \ket{\psi_0} - \braket{\delta\tilde{\psi}_0|\psi_0}\mathcal{E}(\tilde{\psi}_0,\psi_0)\right) = 0,
\end{equation}
due to Eq.~\eqref{eq:constr1} being zero at the solution. Next, we vary the constraint terms, where we note that any tangent vector $\bra{\tilde{u}}$ depends on $\bra{\tilde{\psi}_0}$, and similarly $\ket{u}$ depends on $\ket{\psi_0}$. This dependence can be taken into account by writing $\bra{\tilde{u}} = \bra{\tilde{u}} P_{\tilde{\psi}_0}$ and $\ket{u} = P_{\psi_0}\ket{u}$, respectively, and hence we need to vary only the projection operators.

We are ready to compute the variations of Eq.~\eqref{eq:constr1} and \eqref{eq:constr2} to obtain
\begin{multline}
    \braket{\tilde{\psi}_0|\psi_0}\delta \tilde{L} = \bra{\tilde{u}} \delta P_{\tilde{\psi}_0}(\bar{H}-\mathcal{E}(\tilde{\psi}_0,\psi_0))\ket{\psi_0}   \\
    + \braket{\delta\tilde{\psi}_0|(\bar{H}-\mathcal{E}(\tilde{\psi}_0,\psi_0))|u}, \label{eq:tildepsi0var}
\end{multline}
and
\begin{multline}
    \braket{\tilde{\psi}_0|\psi_0}\delta (\tilde{L}-L) = \bra{\tilde{\psi}_0}(\bar{H}-\mathcal{E}(\tilde{\psi}_0,\psi_0)) \delta P_{\psi_0}\ket{u}   \\
    + \braket{\tilde{u}|(\bar{H}-\mathcal{E}(\tilde{\psi}_0,\psi_0))|\delta{\psi}_0}. \label{eq:psi0var}   
\end{multline}
We have omitted additional terms that vanish at the critical point. We also compute
\[ \delta L = \braket{\phi_0|\Lambda\bar{H}|\delta\psi_0}, \]
which does not vanish in general.


In order to resolve the variations in the projection operators, a small lemma is now useful: Let $P'_{\psi_0}(u)$ be the directional derivative of $P_{\psi_0}$ in the direction $\ket{u}\in T_{\psi_0}\mathcal{M}$. Then, for any $\ket{v} \in T_{\psi_0}\mathcal{M}$,
\begin{subequations}\label{eq:lemma}
\begin{equation}
     P_{\psi_0}'(u)\ket{v} = P_{\psi_0}'(v)\ket{u}, \label{eq:lemma1}
\end{equation}
and by symmetry,
\begin{equation} 
\bra{\tilde{v}} P_{\tilde{\psi}_0}'(\tilde{u}) = \bra{\tilde{u}} P_{\tilde{\psi}_0}'(\tilde{v}). \label{eq:lemma2}
\end{equation}
\end{subequations} 
The proof is as follows: The directional derivative of the projector is
\[  P'_{{\psi}_0}(u) = \sum_{kl} u_l \ket{\partial_l t_k}\bra{t_k} + u_l^* \ket{t_k}\bra{\partial_l t_k}. \]
It is straightforward to show, that any differential of a projection operator $P$ satisfies $\delta P = P \delta P P + (1-P)\delta P (1-P)$, which implies that application of the directional derivative to $\ket{v}\in T_{\psi_0}\mathcal{M}$ eliminates the complex conjugates and gives
\[ \begin{split}
    P'_{\psi_0}(u)\ket{v} &= \sum_{kl} u_l v_k \ket{\partial_l\partial_k \chi(z_0)}  \\& =\sum_{kl} u_l v_k \ket{\partial_k\partial_l \chi(z_0)} = P'_{\psi_0}(v)\ket{u},
    \end{split}
    \]
by symmetry of mixed partial derivatives. We now obtain the useful formula
\[ \delta P_{\psi_0} \ket{v} = P'_{\psi_0}(v)\ket{\delta \psi_0}, \]
and a similar formula for the bra projector variation.

We now assume that the model functions are normalized as $\braket{\tilde{\psi}_0|\psi_0} = 1$, and we set $E = \mathcal{E}(\tilde{\psi}_0,\psi_0)$, the eigenvalue of Eq.~\eqref{eq:cas-evp} and Eq.~\eqref{eq:cas-evp2}. We define $\bar{H}_E = \bar{H}-E$. Equating Eqs.~\eqref{eq:tildepsi0var} and \eqref{eq:psi0var} to zero gives
\begin{subequations}\label{eq:lag-eq-lag2}
\begin{align}
    P_{\tilde{\psi}_0}[P'_{\tilde{\psi}_0}(\tilde{u})\bar{H}_E\ket{\psi_0} + \bar{H}_E\ket{u}] &= 0, \\
    [\bra{\tilde{\psi}_0} \bar{H}_E P'_{\psi_0}(u) + \bra{\tilde{u}} \bar{H}_E + \bra{\phi_0}\Lambda \bar{H}] P_{\psi_0} &= 0,
\end{align}
\end{subequations}
We finally compute the variation in $\tilde{L}$ with respect to $T$, which will give an equation linear in $\Lambda$. We note that $\partial_{t_\mu} \bar{H} = \bar{H}_\mu \equiv [\bar{H},X_\mu]$. Let $\mathcal{F}_\mu(x,y) \equiv \mathcal{E}_{\bar{H}_\mu}(x,y) = \braket{x|\bar{H}_\mu|y}/\braket{x|y}$. It is straightforward to differentiate and get
\begin{multline}
    \partial_{t_\mu} \tilde{L} = \braket{\phi|\Lambda\bar{H}_\mu|\psi_0} + \mathcal{F}_\mu(\tilde{\psi}_0,\psi_0) \\+  \braket{\tilde{u}|\psi_0}(\mathcal{F}_\mu(\tilde{u},\psi_0) - \mathcal{F}_\mu(\tilde{\psi}_0,\psi_0)) \\
    \braket{\tilde{\psi}_0|u}(\mathcal{F}_\mu(\tilde{\psi}_0,u) - \mathcal{F}_\mu(\tilde{\psi}_0,\psi_0)), \label{eq:lambda-eq-lag2}
\end{multline}
which when equated to zero becomes a linear system for $\Lambda$, to be compared with the $\Lambda$ equations for the standard CASCC equations, Eq.~\eqref{eq:lambda-eq}.

Together, Eqs.~\eqref{eq:lag-eq-lag2} and \eqref{eq:lambda-eq-lag2} form a coordinate-independent linear system for $(\ket{u},\bra{\tilde{u}},\Lambda)$. For the sake of concreteness, we expand Eq.~\eqref{eq:lag-eq-lag2} on component form to obtain
\begin{align}
    Au + B\tilde{u} &= 0,\label{eq:linsys1} \\
    A^T\tilde{u} + Cu + D\lambda &= 0,\label{eq:linsys2}
\end{align}
where $A_{kl} = \braket{\tilde{t}_k|\bar{H}_E|t_l}$ and $B_{kl} = \braket{\partial_k \tilde{t}_l|\bar{H}_E|\psi_0}$, a symmetric matrix. Furthermore, $C_{kl} = \braket{\tilde{\psi}_0|\bar{H}_E|\partial_k t_l}$ is symmetric, and $D_{k\mu} = \braket{\phi_\mu|\bar{H}|t_k}$,  and we have collected the $\Lambda$ amplitudes in a vector as $\lambda=(\lambda_\mu)$.

\section{Working equations for the third Lagrangian}

Variation of $\hat{L}$ with respect to $\bra{\tilde{\psi}_0}$ yields
\begin{multline}
    2\delta\hat{L} = \braket{\tilde{v}|\psi_0}^{-1}\bra{\tilde{v}}\delta P_{\tilde{\psi}_0} (\bar{H} - \mathcal{E}(\tilde{v},\psi_0))\ket{\psi_0} \\
    + \braket{\tilde{\psi}_0|v}^{-1}\bra{\delta\tilde{\psi}_0} (\bar{H} - \mathcal{E}(\tilde{v},\psi_0))\ket{v},
\end{multline}
while variation with respect to $\ket{\psi_0}$ yields
\begin{multline}
    2\delta\hat{L} = \braket{\phi_0|\Lambda\bar{H}|\delta\psi_0} +  \braket{\tilde{v}|\psi_0}^{-1}\bra{\tilde{v}} (\bar{H} - \mathcal{E}(\tilde{v},\psi_0))\ket{\delta\psi_0} \\
    + \braket{\tilde{\psi}_0|v}^{-1}\bra{\tilde{\psi}_0} (\bar{H} - \mathcal{E}(\tilde{v},\psi_0))\delta P_{\psi_0} \ket{v}.
\end{multline}
Since $\hat{L}$ is scale invariant with respect to $v$ and $\tilde{v}$, we may choose $\braket{\tilde{v}|\psi_0} = \braket{\tilde{\psi}_0|v} = 1$, and it is readily seen that we obtain a linear system for $v$ and $\tilde{v}$ which is \emph{identical} to that obtained for the second Lagrangian $\tilde{L}$ above, i.e., Eq.~\eqref{eq:lag-eq-lag2}. However, we have not yet determined the equation for $\Lambda$. Differentiation of $\hat{L}$ with respect to $t_\mu$ gives
\begin{equation}
    2\partial_{t_\mu}\hat{L} = \braket{\phi_0|\bar{H}_\mu|\psi_0} +  \mathcal{F}_\mu(\tilde{v},\psi_0) + \mathcal{F}_\mu(\tilde{\psi}_0,v), \label{eq:lambda-eq-lag3}
\end{equation}
which is \emph{not} the same as Eq.~\eqref{eq:lambda-eq-lag2}. Thus, Eq.~\eqref{eq:linsys1}, \eqref{eq:linsys2} (with $u=v$ and $\tilde{u}=\tilde{v}$), and Eq.~\eqref{eq:lambda-eq-lag3} equated to zero form a linear system of equations for the Lagrange multiplier set $(\Lambda,v,\tilde{v})$.

\section{The operator $J$}

Consider a tangent vector $\ket{x}\in T_{\psi_0}\mathcal{M}$ satisfying an equation 
\begin{equation} 
P_{\tilde{\psi}_0}\ket{x} = P_{\tilde{\psi}_0} \ket{y}, 
\label{eq:tangent-eq}
\end{equation}
where $\ket{y}\in\mathcal{H}$ is arbitrary. We desire to solve for $\ket{x}$. Similarly, we also consider a bra tangent vector $\bra{\tilde{x}} \in T_{\tilde{\psi}_0} \mathcal{M}^\dag$ satisfying the 
\begin{equation} 
\bra{\tilde{x}} P_{\psi_0} = \bra{\tilde{y}} P_{\psi_0} .
\label{eq:tangent-eq2}
\end{equation}
Starting with the first equation~\eqref{eq:tangent-eq},
projection onto $\bra{\tilde{t}_k}$ gives the equivalent condition $\braket{\tilde{t}_k|x} = \braket{\tilde{t}_k|y}$. Introduce the overlap matrix $S_{kl} = \braket{\tilde{t}_k|t_l}$, and it is clear that for $\ket{x}$ to be unique, $S$ must be invertible. We expand $\ket{x}$ in the basis for $T_{\psi_0}\mathcal{M}$, i.e., $\ket{x} = \sum_l \ket{t_l} \braket{t_l|x}$. Thus,
\[ \braket{\tilde{t}_k|x} = \sum_l S_{kl} \braket{t_l|x}. \]
Application of $S^{-1}$ from the left gives
\[ \braket{t_l|x} = \sum_{k} S^{-1}_{lk} \braket{\tilde{t}_k|x}. \]
Multiplication with $\ket{t_l}$ from the left and summing gives
\[ \ket{x} = J \ket{y}, \]
where 
\[ J = \sum_{kl} \ket{t_k} S^{-1}_{kl} \bra{\tilde{t}_l}, \]
which therefore acts as a solution operator to Eq.~\eqref{eq:tangent-eq}. A similar argument shows that the operator $J$ also acts as a solution operator for the bra equation~\eqref{eq:tangent-eq2}, i.e., $\bra{\tilde{x}} = \bra{\tilde{y}} J$.

It is straightforward to see that  $J^2 = J$, but in general we have $J^\dag \neq J$, so that $J$ is an oblique projector. The projector happens to be orthogonal whenever $\ket{\psi_0} = \ket{\tilde{\psi}_0}$. We note that for any $\ket{y}\in\mathcal{H}$, $J\ket{y} \in T_{\psi_0}\mathcal{M}$, and for any $\bra{\tilde{y}}\in\mathcal{H}^\dag$, we have $\bra{\tilde{y}}J \in T_{\tilde{\psi}_0}\mathcal{M}^\dag$.  

\section{Equations of motion for the third Lagrangian}

Starting from Eq.~\eqref{eq:action-lag3-prime}, variation of $\psi_0(t)\in\mathcal{M}$ gives
\begin{equation}
    \begin{split}    
    2\delta \hat{S} &= \int_{t_0}^{t_1} i \braket{\phi_0|\Lambda\dot{T}|\delta\psi_0} - i \braket{\dot{\tilde{v}}|\delta{\psi_0}} - i \braket{\dot{\tilde{\psi}}_0|\delta P_{\psi_0}|v} \\ &\qquad - i \braket{\tilde{v}|\dot{P}_{\tilde{\psi}_0}|\delta\psi_0} - 2\delta\hat{\mathcal{K}} \, dt \\
    &= \int_{t_0}^{t_1} i \braket{\phi_0|\Lambda\dot{T}|\delta\psi_0} - i \braket{\dot{\tilde{v}}|\delta{\psi_0}} - i \braket{\dot{\tilde{\psi}}_0|P'_{\psi_0}(v)|\delta\psi_0} \\ &\qquad - i \braket{\dot{\tilde{\psi}}_0|P'_{\tilde{\psi}_0}(\tilde{v})|\delta\psi_0} - 2\delta\hat{\mathcal{K}} \, dt
    \end{split}
\end{equation}
Further simplification reads
\begin{equation}
    \begin{split}
    2 \delta\hat{S} &= \int_{t_0}^{t_1} i \braket{\phi_0|\Lambda\dot{T}|\delta\psi_0} - i \braket{\dot{\tilde{v}}|\delta{\psi_0}} 
    \\ &\qquad + \braket{\tilde{\psi}_0|\bar{H} J (P'_{\tilde{\psi}_0}(\tilde{v}) + P'_{\psi_0}(v))|\delta\psi_0} - 2\delta\hat{\mathcal{K}} \, dt
    \end{split}
\end{equation}
Here, we used Eqs.~\eqref{eq:lemma} and the equation of motion~\eqref{eq:left-tdse}. The variation in $\hat{\mathcal{K}}$ is
\begin{equation}
    2 \delta \hat{\mathcal{K}} = \braket{\phi_0|\Lambda\bar{H}|\delta\psi_0} + \bra{\tilde{v}}\bar{H}\ket{\delta\psi_0} + \bra{\tilde{\psi}_0}\bar{H} P'_{\psi_0}(v)\ket{\delta \psi_0} .
\end{equation}
Since $\delta\hat{S}=0$ is required for arbitrary $\delta\psi(t)\in T_{\psi_0}\mathcal{M}$, we obtain the Euler--Lagrange equations
\begin{subequations} \label{eq:multiplier-eom-lag3-deriv}
\begin{equation}
    \begin{split}
        -i \bra{\dot{\tilde{v}}} &=  \bra{\phi_0}\Lambda(\bar{H}-i\dot{T})J + \bra{\tilde{v}} \bar{H} J \\ &\quad -  \bra{\tilde{\psi}_0} \bar{H} (J P'_{\tilde{\psi}_0}(\tilde{v}) + [J,P'_{\psi_0}(v)] ) J     .
    \end{split} 
\end{equation}
A symmetric argument, starting from Eq.~\eqref{eq:action-lag3}, gives the equation of motion
\begin{equation}
    \begin{split}
        i \ket{\dot{v}} &=  J \bar{H}\ket{v}  -  J ( P'_{\psi_0}(v) J + [P'_{\tilde{\psi}_0}(\tilde{v}),J]) \bar{H} \ket{\psi_0}.
    \end{split}
\end{equation}
\end{subequations}


\bibliographystyle{unsrt}
\bibliography{refs}

\end{document}